\documentclass[10pt,a4paper,twocolumn,english,twocolumn]{revtex4}
\usepackage[T1]{fontenc}
\usepackage[latin1]{inputenc}
\usepackage{graphicx}

\makeatletter

\newcommand{\noun}[1]{\textsc{#1}}

\providecommand{\tabularnewline}{\\}

\usepackage{babel}
\makeatother
\begin{document}

\title{Constraining the growth factor with baryon oscillations}

\author{Domenico Sapone$^{1}$}

\email{domenico.sapone@physics.unige.ch}

\author{Luca Amendola$^{2}$}

\email{amendola@mporzio.astro.it}

\address{$^{1}$Départment de Physique Théorique, Université de Genève, 24
quai Ernest Ansermet, CH-1211, Genève 4, Switzerland }

\address{$^{2}$INAF/Osservatorio Astronomico di Roma Via Frascati 33 - 00040
Monteporzio Catone (Roma) - Italy}

\begin{abstract}
The growth factor of linear fluctuations is probably one of the least
known quantity in observational cosmology. Here we discuss the contraints
that baryon oscillations in galaxy power spectra from future surveys
can put on a conveniently parametrized growth factor. We find that
spectroscopic surveys of 5000 deg$^{2}$ extending to $z\approx3$
could estimate the growth index $\gamma$ within $0.06$; a similar
photometric survey would give $\Delta\gamma\approx0.15$. This test
provides an important consistency check for the standard cosmological
model and could constrain modified gravity models. We discuss the
errors and the figure of merit for various combinations of redshift
errors and survey sizes.
\end{abstract}
\maketitle

\section{Introduction}

The characterization of dark energy (DE) has been so far based almost
uniquely on background tests at rather low redshifts ($z\le1.5$:
Riess et al. 1998, Perlmutter et al. 1999, Tonry et al. 2003, Riess
et al. 2004, Astier et al. 2006, Eisenstein et al. 2005) or very large
redshifts ($z\approx1000$: e.g. Netterfield et al. 2002, Halverson
et al. 2002, Lee et al. 2002, Bennett et al. 2003, Spergel et al.
2006). These tests are based essentially on estimations of luminosity
$D_{L}\left(z\right)$ or angular-diameter distances $D(z)$, i.e.
on integrals of the Hubble function $H(z)$ which, in turn, contain
integrals of the equation of state. Only very recently tests involving
the linear perturbations have begun to be discussed, using methods
based on the integrated Sachs-Wolfe effect, weak lensing and high-redshift
power spectra (e.g. Boughn \& Crittenden 2004, Refregier et al. 2006,
Crotts et al. 2005 ). However, it is fair to say that the growth function
is still one of the least known quantity in cosmology. So far, it
is possible to quote only two published results that put limits on
it: the value at $z\approx0.15$ obtained in 2dF (Hawkins et al. 2003;
Verde et al. 2002) and the $z\approx3$ result from Lyman$-\alpha$
clouds (McDonald et al. 2005). Defining $G(z)=\delta(z)/\delta(0)$
($\delta$ being the matter density contrast) we have for \begin{equation}
f\equiv\frac{d\log G}{d\log a}\label{eq:growthindex}\end{equation}
the value $f=0.51\pm0.15$ for 2dF at $z\approx0.15$ and $f=1.46\pm0.29$
for the Lyman-$\alpha$ at $z\approx3$ . These results show clearly
how large is the degree of uncertainty. Actually the uncertainty is
much larger than it appears from the quoted statistical errors. In
the case of the low-$z$ estimate, the result is obtained by estimating
the bias from higher-order statistics, which is known to be particularly
sensitive to the selection effects, to incompleteness etc.; different
methods give in fact quite different results (see discussion in Hawkins
et al. 2003). In the case of the high-$z$ estimation, the main problem
is the reconstruction of the bias factor from numerical simulations
which, by their nature, are performed only in a limited range of fiducial
models. It is therefore important to test the growth factor with other
methods and with improved datasets. 

A test of the growth factor would be important both as a consistency
check for the standard cosmological model (since $f$ is determined
by $H(z)$ in a standard cosmology) and as a constraint on non-standard
models like e.g. modified gravity. In fact, models that modify the
Poisson equation will also generically modify the perturbation equation
for the matter density contrast $\delta$. As an example, models in
which dark energy is coupled to matter display a growth index which
deviates from the standard case at all epochs (see e.g. Amendola \&
Tocchini-Valentini 2003; Demianski et al. 2004; Nunes \& Mota 2004).
Several other papers discussed the parametrization of the perturbation
equations in modified gravity models, see e g. Ishak et al. (2005),
Heavens, Kitching and Taylor (2006), Taylor et al. (2007); Heavens,
Kitching, Verde (2007) , Caldwell, Cooray and Melchiorri (2007), Amendola,
Kunz and Sapone (2007), Zhang et al. (2007).

In this paper we investigate the extent to which baryon oscillations
can set limits to $G(z)$ in future large-scale observations at $z$
up to 3. The method we use is based on recent proposals (Linder 2003,
Blake \& Glazebrook 2003, Seo \& Eisenstein 2003) to exploit the baryon
acoustic oscillations (BAOs) in the power spectrum as a standard ruler
calibrated through CMB acoustic peaks. In particular, Seo \& Eisenstein
(2003; SE) have shown the feasibility of large (100 to 1000 square
degrees) spectroscopic surveys at $z\approx1$ and $z\approx3$ to
put stringent limits to the equation of state $w(z)$ and its derivative.
As it is well-known, BAOs have been detected at low $z$ in SDSS (Eisenstein
et al. 2005); the detection at large $z$ , where more peaks at smaller
scales can be obtained, is likely to become one of the most interesting
astrophysical endeavours of the next years.

\section{Background equation}

Here we review the basic equations and notation for the background
evolution and for the linear fluctuations. The evolution of the dark
energy can be expressed by the present dark energy density $\Omega_{DE}$
and by a time-varying equation of state (see Copeland, Sami \& Tsujikawa
2006 for a recent review): \begin{equation}
w(z)=\frac{p}{\rho}\label{eq:parametro di stato}\end{equation}

Given $w(z)$, the dark energy density equation is $\rho(z)=\rho(0)a^{-3(1+\hat{w})}$
where \begin{equation}
\hat{w}(z)=\frac{1}{\log(1+z)}\int_{0}^{z}\frac{w(z')}{1+z'}dz'\label{eq:densit}\end{equation}

The Hubble parameter and the angular diameter distance,  $H(z)$ and
$D_{A}(z)$, assuming a flat universe $\Omega_{m}+\Omega_{DE}=1$,
 become respectively: \begin{equation}
H^{2}\left(z\right)=H_{0}^{2}[\Omega_{m_{0}}(1+z)^{3}+(1-\Omega_{m_{0}})(1+z)^{3(1+\hat{w})}]\label{eq:parametro-di-Hubble}\end{equation}
and \begin{equation}
D_{A}(z)=\frac{c}{1+z}\int_{0}^{z}\frac{dz'}{H(z')}\label{eq:distanza-angolare}\end{equation}
where the total matter density is\begin{equation}
\Omega_{m}(z)=\frac{\Omega_{m_{0}}}{\Omega_{m_{0}}+(1-\Omega_{m_{0}})(1+z)^{3\hat{w}}}\label{eq:omgen}\end{equation}

It is well known that a good approximation to the growth index for
sub-horizon scales in flat models is given by (Lahav et al. 1991,
Wang and Steinhardt 1998 ) \begin{equation}
f\equiv\frac{\partial\log G}{\partial\log a}=\Omega_{m}\left(a\right)^{\gamma}\label{eq:growthindex1}\end{equation}
This introduces a new parameter $\gamma$, beside those that characterize
the background model (see also Linder 2005, Percival 2005) .

We remark that a recent analysis of most of the extant data produced
the result $\gamma=0.6_{-0.3}^{+0.4}$ (Di Porto \& Amendola 2007)
.

\section{Fisher matrix formalism}

Following Seo \& Eisenstein (2003; hereinafter SE) we write schematically
the observed galaxy power spectrum as:\begin{eqnarray}
P_{obs}(z,k_{r}) & = & \frac{D_{Ar}^{2}(z)H(z)}{D_{A}^{2}(z)H_{r}(z)}G^{2}(z)b(z)^{2}\left(1+\beta\mu^{2}\right)^{2}P_{0r}(k)\nonumber \\
 &  & +P_{shot}(z)\end{eqnarray}
where the subscript $r$ refers to the values assumed for the reference
cosmological model, i.e. the model at which we evaluate the Fisher
matrix. Here $P_{shot}$ is the shot noise due to discreteness in
the survey, $\mu$ is the direction cosine within the survey, $P_{0}$
is the present spectrum for the fiducial cosmology. For the linear
matter power spectrum we adopt the fit by Eisenstein \& Hu (1999)
(with no massive neutrinos and also neglecting any change of the shape
of the spectrum from small deviation around $w=-1$).

The wavenumber $k$ is also to be transformed between the fiducial
cosmology and the general one (SE; see also Amendola, Quercellini,
Giallongo 2004, hereinafter AQG, for more details). The bias factor
is defined as: \begin{equation}
b(z)=\frac{\Omega_{m}\left(z\right)^{\gamma}}{\beta(z)}\label{eq:bias}\end{equation}
and for the fiducial model is estimated by comparing the $8$Mpc$/h$
cell variance $\sigma_{8,g}$ of the galaxies corrected for the linear
redshift distortion with the same quantity for the total matter. Clearly,
the growth function is degenerate with the bias except for the redshift
correction factor $(1+\beta\mu^{2})$. Since we marginalize over $\beta$,
it is clear that the redshift correction plays a crucial role for
as concern the estimation of the growth factor. The linear correction
we use should therefore be considered only a first approximation and
more work to go beyond Kaiser's small-angle and Gaussian approximation
is needed, as discussed in Hamilton \& Culhane (1996), Zaroubi \&
Hoffman (1996), Tegmark et al. (2004) and Scoccimarro (2004).

The total galaxy power spectrum including the errors on redshift can
be written as (SE) \begin{equation}
P\left(z,k\right)=P_{obs}\left(z,k\right)e^{k^{2}\mu^{2}\sigma_{r}^{2}}\label{eq:ps}\end{equation}
where $\sigma_{r}=\frac{\delta z}{H\left(z\right)}$ is the absolute
error on the measurement of the distance and $\delta z$ is the absolute
error on redshift. Given the uncertainties of our observations, we
now want to propagate these errors to compute the constraints on cosmological
parameters. The Fisher matrix provides a useful method for doing this.
Assuming the likelihood function to be Gaussian, the Fisher matrix
is (Eisenstein, Hu \& Tegmark 1998; Tegmark 1997) \begin{equation}
F_{ij}=2\pi\int_{k_{min}}^{k_{max}}\frac{\partial\log P\left(k_{n}\right)}{\partial\theta_{i}}\frac{\partial\log P\left(k_{n}\right)}{\partial\theta_{j}}\cdot V_{eff}\cdot\frac{k^{2}}{8\pi^{3}}\cdot dk\label{eq:FisherMatrix}\end{equation}
where the derivatives are evaluated at the parameter values of the
fiducial model and $V_{eff}$ is the effective volume of the survey,
given by:\begin{eqnarray}
V_{eff} & = & \int\left[\frac{n\left(\vec{r}\right)P\left(k,\mu\right)}{n\left(\vec{r}\right)P\left(k,\mu\right)+1}\right]^{2}d\vec{r}=\nonumber \\
 & = & \left[\frac{n\left(\vec{r}\right)P\left(k,\mu\right)}{n\left(\vec{r}\right)P\left(k,\mu\right)+1}\right]^{2}V_{survey}\label{eq:Volume}\end{eqnarray}
where the last equality holds only if the comoving number density
is constant in position and where $\mu=\vec{k}\cdot\widehat{r}/k$,
$\widehat{r}$ being the unit vector along the line of sight and $k$
the wave vector. The highest frequency $k_{max}(z)$ is chosen to
be near the scale of non-linearity at $z$: we choose values from
$0.11h$/Mpc for small $z$ bins to $0.33h$/Mpc for the highest redshift
bins. Any submatrix of $F_{ij}^{-1}$ gives the correlation matrix
for the parameters corresponding to rows and columns on that submatrix.
The eigenvectors and eigenvalues of this correlation matrix give the
orientation and the size of the semiaxes of the confidence region
ellipsoid. This automatically marginalizes over the remaining parameters.
The parameters that we use for evaluating the Fisher matrix are shown
in Tab. (\ref{tab:Cosmological-parameters}). Our fiducial model corresponds
to the $\Lambda$CDM WMAP3y best-fit parameters (Spergel et al. 2006):
$\Omega_{m0}=0.28$, $h=0.73$, $\Omega_{DE}=0.72$, $\Omega_{K}=0$,
$\Omega_{b}h^{2}=0.0223$, $\tau=0.092$, $n_{s}=0.96$ and $T/S=0$
and as anticipated $\gamma=0.545$. Beside the BAO from large scale
structure, we also employ the CMB Fisher matrix, following the method
in Eisenstein, Hu \& Tegmark (1999) and assuming a Planck-like experiment.
The cosmological parameters we use for CMB are listed in Tab. (\ref{tab:CMB-parameters}).
The total Fisher matrix is given simply by the addition of the two
matrices.

\begin{table}
\begin{centering}\begin{tabular}{|c|c|c|}
\hline 
&
\textbf{Parameters}&
\tabularnewline
\hline 
1&
total matter density&
$\omega_{m}=\Omega_{m_{0}}h^{2}$\tabularnewline
\hline 
2&
total baryon density&
$\omega_{b}=\Omega_{b_{0}}h^{2}$\tabularnewline
\hline 
3&
optical thickness&
$\tau$\tabularnewline
\hline 
4&
spectral index&
$n_{s}$\tabularnewline
\hline 
5&
present matter density &
$\Omega_{m_{0}}$\tabularnewline
\hline 
&
&
\tabularnewline
\hline 
&
\emph{For each redshift bin}&
\tabularnewline
\hline 
&
&
\tabularnewline
\hline 
6&
shot noise&
$P_{s}$\tabularnewline
\hline 
7&
angular diameter distance&
$\log D_{A}$\tabularnewline
\hline 
8&
Hubble parameter&
$\log H$\tabularnewline
\hline 
9&
growth factor&
$\log D$\tabularnewline
\hline 
10&
bias&
$\log\beta$\tabularnewline
\hline
\end{tabular}\par\end{centering}

\caption{Cosmological parameters\label{tab:Cosmological-parameters}}
\end{table}

The derivatives of the spectrum with respect to the cosmological parameters
$p_{i}$ (i.e. $\omega_{m}=\Omega_{m_{0}}h^{2}$, $\omega_{b}=\Omega_{b_{0}}h^{2}$,
$\tau$, $n_{s}$
, $\Omega_{m_{0}}$ plus $P_{s},\beta,G,D,H$ for each redshift bin)
are evaluated using the fit of Eisenstein \& Hu (1999).

\begin{table}
\begin{centering}\begin{tabular}{|c|c|c|}
\hline 
&
\textbf{Parameters}&
\tabularnewline
\hline 
1&
total matter density&
$\omega_{m}=\Omega_{m_{0}}h^{2}$\tabularnewline
\hline 
2&
total baryon density&
$\omega_{b}=\Omega_{b}h^{2}$\tabularnewline
\hline 
3&
optical thickeness&
$\tau$\tabularnewline
\hline 
4&
spectral index&
$n_{s}$\tabularnewline
\hline 
5&
matter density today&
$\Omega_{m_{0}}$\tabularnewline
\hline 
6&
tensor scalar ratio&
$T/S$\tabularnewline
\hline 
7&
angular diameter distance&
$\log D_{A}$\tabularnewline
\hline 
8&
normalization factor&
$\log A_{s}$\tabularnewline
\hline
\end{tabular}\par\end{centering}

\caption{CMB parameters\label{tab:CMB-parameters}}
\end{table}

Since we want to propagate the errors to the cosmologically relevant
set of parameters\begin{equation}
q_{i}=\left\{ \, w_{0},\, w_{1},\gamma\right\} \label{eq:q_i}\end{equation}
 we need to change parameter space. This will be done taking the inverse
of the Fisher Matrix $F_{ij}^{-1}$ and then extracting a submatrix,
called $F_{mn}^{-1}$ containing only the rows and columns with the
parameters that depend on $q_{i}$, namely $D_{A}$, $H$ and $G$.
The root mean square of the diagonal elements of the inverse of the
submatrix give the errors on $D_{A}$, $H$, and $G$. Then we contract
the inverse of the submatrix with the new set of parameters $q_{i}$;
the new Fisher matrix will be given by \begin{equation}
\overline{F}_{DE;ij}=\frac{\partial p_{m}}{\partial q_{i}}\,\overline{F}_{mn}\,\frac{\partial p_{n}}{\partial q_{j}}\label{eq:subfisher}\end{equation}
This automatically marginalizes over all the remaining parameters.

The derivatives of the Hubble parameter and for the angular diameter
distance can be written as \begin{equation}
\frac{\partial\log~H}{\partial q_{i}}=\frac{1}{H}\frac{\partial H}{\partial q_{i}}\label{eq:dlogHdq}\end{equation}

\begin{equation}
\frac{\partial\log~D_{A}}{\partial q_{i}}=-\frac{1}{\left(1+z\right)D_{A}}\int\frac{\partial\log~H}{\partial q_{i}}\frac{1}{H}dz\label{eq:dlogDdq}\end{equation}

\section{growth factor}

\begin{figure}
\begin{centering}\includegraphics[scale=0.3]{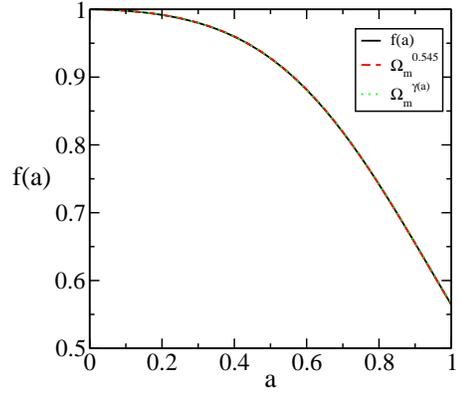}\par\end{centering}

\caption{\label{fig:Growth-index-DE}Growth index vs the scale factor for
a DE model with a varying equation of state, $w(z)=w_{0}+w_{1}z$
with $w_{0}=-1.5$ and $w_{1}=1$. The black solid line refers to
the solution obtained by the differential equation for $\gamma$ (Percival
2005). The red dashed line refers to the growth index given by eq.
(\ref{eq:growthindex1}) and the green dotted line is the growth index
with a $\gamma$ factor given by eq. (\ref{eq:gamma}). The matter
density is given by eq. (\ref{eq:omgen}).}
\end{figure}

We consider now separately two cases: in Case 1 the growth rate depends
on $w$ (assumed constant); in Case 2 the growth rate is free and
we forecast the constraints that future experiments can put on it.

\begin{table}
\begin{centering}\begin{tabular}{|c|c|c|}
\hline 
\multicolumn{3}{|c|}{Surveys}\tabularnewline
\hline
$z$&
 $V_{s}$ (Gpc/h)$^{3}$&
 $n$\tabularnewline
\hline
$0-0.5$&
 0.006&
 $5\cdot10^{-2}$\tabularnewline
\hline
$0.5-0.7$&
 0.0082&
 $6.9\cdot10^{-2}$\tabularnewline
\hline
$0.7-0.9$&
 0.011&
 $4.2\cdot10^{-2}$\tabularnewline
\hline
$0.9-1.1$&
 0.0135&
 $3.1\cdot10^{-2}$\tabularnewline
\hline
$1.1-1.3$&
 0.015&
 $2.4\cdot10^{-2}$\tabularnewline
\hline
$2.7-3.5$&
 0.073&
 $2\cdot10^{-3}$\tabularnewline
\hline
\end{tabular}\par\end{centering}

\caption{\label{tab:surveys}Details of the surveys.}
\end{table}

\begin{figure}
\begin{centering}\includegraphics[scale=0.51]{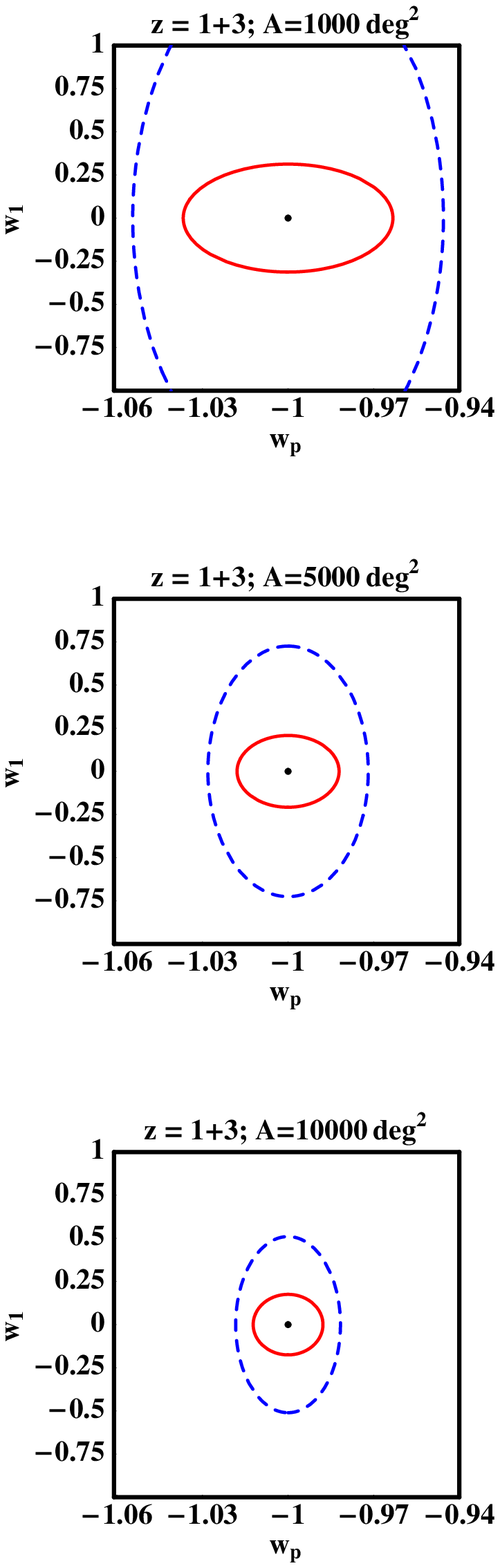}\includegraphics[scale=0.5]{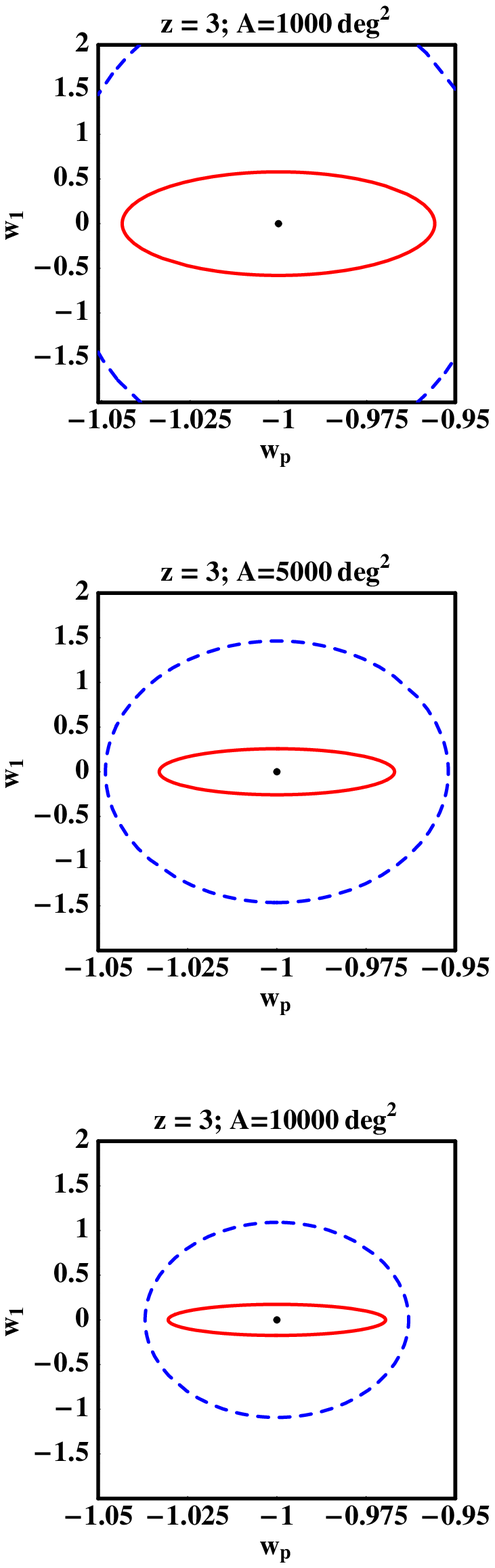}\par\end{centering}

\caption{\label{fig:w0-w1-2}Confidence level for $w_{p}$ and $w_{1}$ for
surveys of 1000, 5000 and 10000 $deg^{2}$ and for different combinations
of redshift bins (case 1). The solid red curve refers to spectroscopic
surveys and the dashed blue curve to photometric surveys, $\delta z=0$
and $\delta z/z=0.04$ respectively.}
\end{figure}

\begin{figure}
\begin{centering}\includegraphics[scale=0.5]{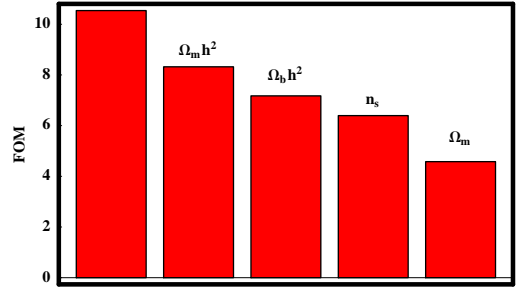}\par\end{centering}

\caption{\label{fig:foms-marginalizing}Fom for $w_{p}-w_{1}$ vs marginalized
parameters. }
\end{figure}

\subsection{Case 1}

In general, the exponent $\gamma$ depends on the cosmological parameters.
To see this, we just need to consider the equation of perturbations
and insert the growth index defined by eq. (\ref{eq:growthindex}).
Then we obtain the approximate analytic solution (Wang \& Steinhardt
1998) : 

\begin{equation}
\gamma\left(z\right)=\frac{3}{5-\frac{w\left(z\right)}{1-w\left(z\right)}}\label{eq:gamma}\end{equation}
and for a $\Lambda$CDM model $\gamma=0.545$. The behavior of the
growth index for a $w(z)$ model is shown in Fig. (\ref{fig:Growth-index-DE}).
We can see that there is almost no difference in behavior between
the curves obtained with the approximation (\ref{eq:gamma}). Because
of the dependence of $\gamma$ on the dark energy parameters, the
derivatives of the growth factor are given by: \begin{eqnarray}
\frac{\partial\log~G}{\partial q_{i}} & = & -\int\left[\frac{\partial\gamma}{\partial q_{i}}\log\Omega_{m}(z)\right.\nonumber \\
 &  & \left.+\gamma\frac{\partial{\log\Omega_{m}(z)}}{\partial q_{i}}\right]\Omega_{m}(z)^{\gamma}\frac{dz}{\left(1+z\right)}\label{eq:dlogGdq}\end{eqnarray}

In this case the new set of parameters is $q_{i}=\left\{ \, w_{0},\, w_{1}\right\} $
and we assume as fiducial model $w_{0}=-1,w_{1}=0$. The factor $\gamma$,
in this case, depends only on the dark energy parameters $w_{0}$
and $w_{1}$; this means the only non-vanishing derivatives are $\frac{\partial\gamma}{\partial w_{0}}$
and $\frac{\partial\gamma}{\partial w_{1}}$. In Fig. (\ref{fig:w0-w1-2})
the confidence regions are shown for different combination of redshift
and area. Instead of $(w_{0},w_{1})$ we use the pivot parameters
$w_{p}-w_{1}$ (projection of $w_{0}-w_{1}$ on the pivot point, defined
as the value of $z$ for which the uncertainty in $w\left(z\right)$
is smallest).

\begin{figure}
\begin{centering}\includegraphics[scale=0.5]{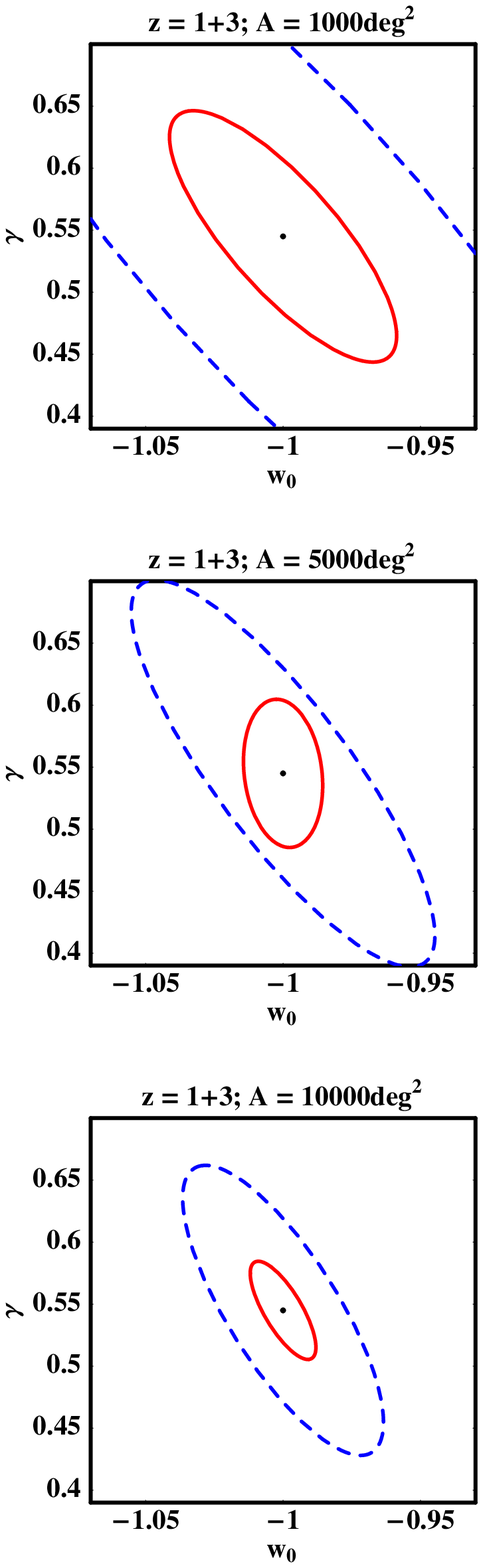}\includegraphics[scale=0.5]{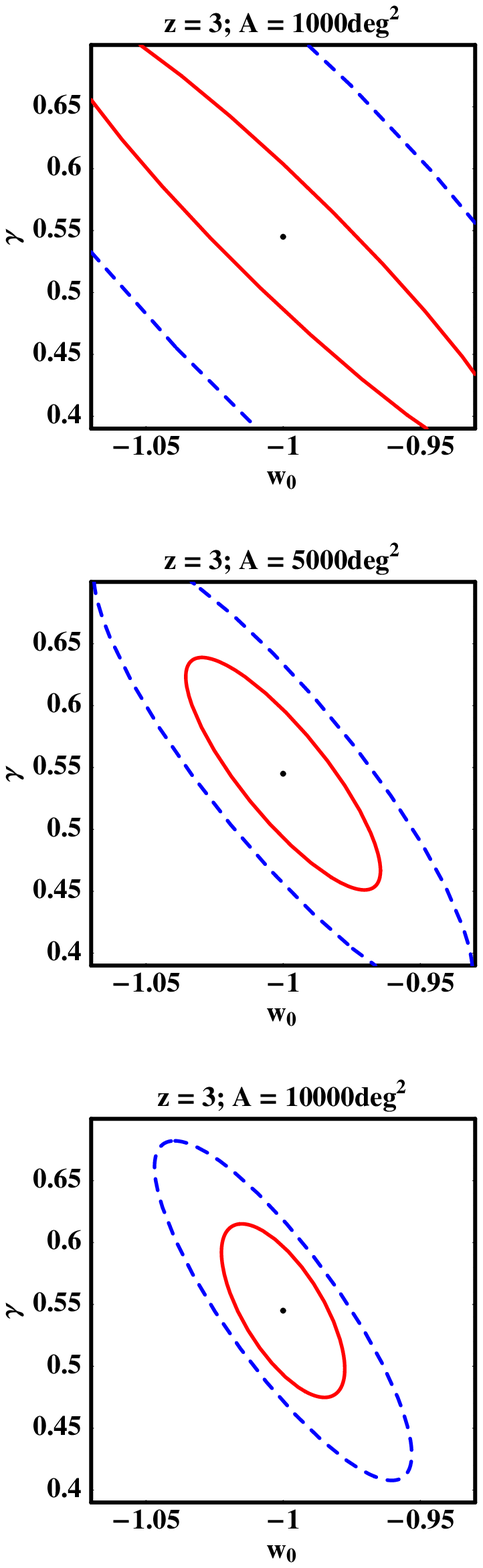}\par\end{centering}

\caption{\label{fig:w0-gamma-2}Confidence level for $w_{0}$ and $\gamma$
for surveys of 1000, 5000 and 10000 $deg^{2}$ and for different combinations
of redshift bins (case 2). The solid red curve refers to spectroscopic
surveys and the dashed blue curve to photometric surveys, $\delta z=0$
and $\delta z/z=0.04$ respectively.}
\end{figure}

\begin{table}
\begin{centering}\begin{tabular}{|c|c|c|c|c|c|c|}
\hline 
 &
\multicolumn{2}{c|}{$1000\: deg^{2}$}&
\multicolumn{2}{c|}{$5000\: deg^{2}$}&
\multicolumn{2}{c|}{$10000\: deg^{2}$}\tabularnewline
\hline 
$\delta z$&
$w_{p}$&
$w_{1}$&
$w_{p}$&
$w_{1}$&
$w_{p}$&
$w_{1}$\tabularnewline
\hline 
&
\multicolumn{6}{c|}{$z=1+3$}\tabularnewline
\hline 
$0$\%&
$0.036$&
$0.313$&
$0.018$&
$0.208$&
$0.012$&
$0.175$\tabularnewline
\hline 
$4$\%&
$0.054$&
$1.523$&
$0.028$&
$0.726$&
$0.018$&
$0.51$\tabularnewline
\hline 
&
\multicolumn{6}{c|}{$z=3$}\tabularnewline
\hline 
$0$\%&
$0.044$&
$0.579$&
$0.033$&
$0.257$&
$0.030$&
$0.154$\tabularnewline
\hline 
$4$\%&
$0.061$&
$2.614$&
$0.051$&
$1.463$&
$0.034$&
$1.092$\tabularnewline
\hline
\end{tabular}\par\end{centering}

\caption{\label{tab:w0-w1-2} Values of $\sigma_{w_{p}}$ and $\sigma_{w_{1}}$
for spectroscopic surveys $\delta z=0$ and photometric surveys $\delta z=4$\%
on the redshift estimate and for several survey areas (case 1). We
consider two different combinations of redshift ($z=1+3$ and $z=3$
only).}
\end{table}

\subsection{Case 2}

We want now to put constraints on $\gamma$ as a free parameter. We
assume here $w=constant$ and again $w_{0}=-1$ as fiducial value.
The new set of parameters is therefore $q_{i}=\left\{ \, w_{0},\,\gamma\right\} $.
The derivatives with respect to the first three parameters are given
by the eq. (\ref{eq:dlogGdq}). The derivative for the growth factor
with respect to $\gamma$ is: \begin{eqnarray}
\frac{\partial\log G}{\partial\gamma} & = & -\int\frac{\partial}{\partial\gamma}\exp\left[\gamma\log\Omega_{m}\left(z\right)\right]\frac{dz}{\left(1+z\right)}=\nonumber \\
 & = & -\int\log\Omega_{m}\left(z\right)\,\Omega_{m}\left(z\right)^{\gamma}\frac{dz}{\left(1+z\right)}\label{eq:dlogGdgamma}\end{eqnarray}

\begin{table}
\begin{centering}\begin{tabular}{|c|c|c|c|c|c|c|}
\hline 
 &
\multicolumn{2}{c|}{$1000\: deg^{2}$}&
\multicolumn{2}{c|}{$5000\: deg^{2}$}&
\multicolumn{2}{c|}{$10000\: deg^{2}$}\tabularnewline
\hline 
$\delta z$&
$w_{0}$&
$\gamma$&
$w_{0}$&
$\gamma$&
$w_{0}$&
$\gamma$\tabularnewline
\hline 
&
\multicolumn{6}{c|}{$z=1+3$}\tabularnewline
\hline 
$0$\%&
$0.045$&
$0.099$&
$0.016$&
$0.059$&
$0.004$&
$0.05$\tabularnewline
\hline 
$4$\%&
$0.128$&
$0.301$&
$0.062$&
$0.153$&
$0.044$&
$0.114$\tabularnewline
\hline 
&
\multicolumn{6}{c|}{$z=3$}\tabularnewline
\hline 
$0$\%&
$0.089$&
$0.188$&
$0.039$&
$0.092$&
$0.026$&
$0.069$\tabularnewline
\hline 
$4$\%&
$0.152$&
$0.344$&
$0.076$&
$0.18$&
$0.081$&
$0.197$\tabularnewline
\hline
\end{tabular}\par\end{centering}

\caption{\label{tab:w0-gamma-2}Values of $\sigma_{w_{0}}$ and $\sigma_{\gamma}$
for spectroscopic surveys $\delta z=0$ and photometric surveys $\delta z/z=4$\%
on the measure of the redshift and for several areas (case 2). We
consider two different combinations of redshift bins ($z=1+3$ and
$z=3$ only). }
\end{table}

\begin{figure}
\begin{centering}\includegraphics[scale=0.53]{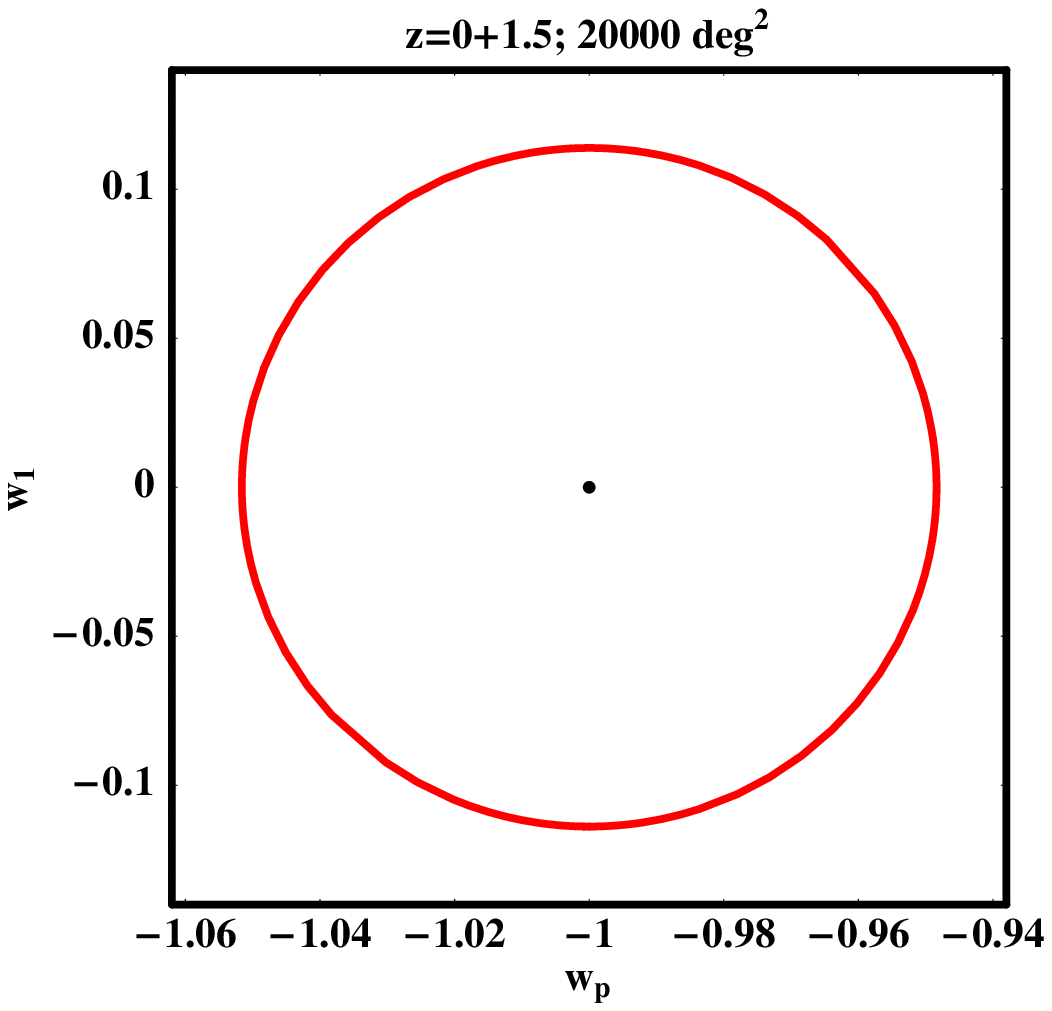}\par\end{centering}

\caption{\label{fig-DETF-wp-w1}Confidence level for $w_{p}$ and $w_{1}$
for surveys 20000 $deg^{2}$ (DETF case). The solid red curve refers
to spectroscopic surveys.}
\end{figure}

\begin{figure}
\begin{centering}\includegraphics[scale=0.5]{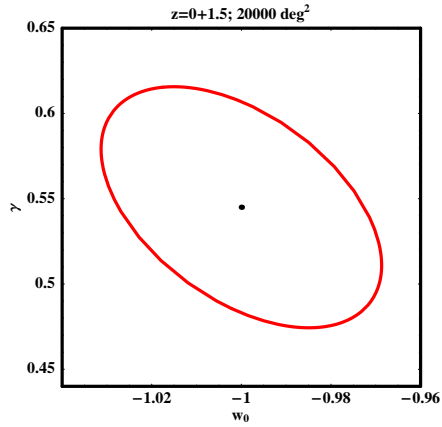}\par\end{centering}

\caption{\label{fig-DETF-wp-gamma}Confidence level for $w_{0}$ and $\gamma$
for surveys 20000 $deg^{2}$ (DETF case). The solid red curve refers
to spectroscopic surveys.}
\end{figure}

\section{Results and Conclusions}

The main aim of this work is to give marginalized constraints on the
dark energy parameters ($w_{p}-w_{1}$) and most importantly on the
growth factor itself, for several combinations of surveys, redshift
errors and area. Following SE and Amendola, Quercellini, Giallongo
(2004) we consider several binned surveys with average redshift depth
around $z=1$ and $z=3$ plus a SDSS-like survey at $z<0.5$, as detailed
in Table \ref{tab:surveys}. More details can be found in AQG. We
consider both spectroscopic surveys ($\delta z=0$ ) and photometric
surveys ($\delta z/z=0.04$) and three areas ($1000,\,5000,\,10000$
deg$^{2}$). These features are well within the range of proposed
experiments like JDEM and DUNE (Crotts at al. 2005; Réfrégier et al.
2006; see also DETF Report Albrecht et al. 2006) 

We first consider Case I, in which the growth factor is not an independent
quantity but is a function of $w(z)$. The two-dimensional regions
of confidence are shown in Fig (\ref{fig:w0-w1-2}) and the final
marginalized errors are summarized in Tab. (\ref{tab:w0-w1-2}). The
errors on $w_{p}$ reduce from 0.036 to 0.012 for the spectroscopic
case for surveys that extend from 1000 to 10000 deg$^{2}$ and from
0.054 to 0.018 in the photometric case.

Then we consider Case II, in which $\gamma$ is a free constant as
in eq. (\ref{eq:growthindex}). In Fig. (\ref{fig:w0-gamma-2}) we
show the confidence regions for $w_{0}-\gamma$. The errors are given
in Tab. (\ref{tab:w0-gamma-2}) . We see that the errors on $\gamma$
reduce from 0.099 to 0.05 for spectroscopic surveys and from 0.301
to 0.114 in the photo-$z$ case. These errors are way too large to
produce an independent constraint on $w$ (in fact one has approximately
$\Delta w\approx10\Delta\gamma$ near $w=-1$) but, beside being a
general test of consistency for the cosmological model, they would
certainly give interesting constraints on models that predict growths
different from standard, like modified gravity models (Koyama \& Maartens
2005; Maartens 2006; Amendola, Charmousis \& Davis 2005; Amendola,
Polarski, Tsujikawa 2006). In Fig. (\ref{fig:foms-marginalizing})
we show the FOM for $w_{p}-w_{1}$, for only one survey ($5000\, deg^{2}$)
and only one combination of redshift ($z=1-3$), first when all the
other parameters are fixed and then successively marginalizing over
the parameter indicated and over all those on the left (eg the third
column represents the marginalization over $\omega_{m}$, $\omega_{b}$).

We can compare our results to those obtained recently by Huterer and
Linder (2006). Using a combination of weak lensing, SNIa and CMB methods,
they predict $\sigma\left(\gamma\right)=0.044$ for future experiments.
With large-scale tomographic weak lensing alone, Amendola, Kunz, and
Sapone (2007) predict $\sigma\left(\gamma\right)=0.04$ at $68\%$
confidence level. These values are comparable to those  obtained here
with the BAO method and considering a spectroscopic survey of 5000$deg^{2}$,
$\sigma\left(\gamma\right)=0.059$. 

We notice that the difference on the growth index $\gamma$ between
General Relativity and an extradimensional gravity model (as DGP,
where $\gamma=0.68$, see Linder \& Cahn 2007) is $\Delta\gamma=0.135$;
if we compare our results shown in Tab. (\ref{tab:w0-gamma-2}) we
see that the errors on $\gamma$ for a photometric survey are within
this range, meaning that DGP model cannot be escluded. Things get
slightly better if we consider spectroscopic surveys, where errors
decrease with about $30\%$; however in this case we require a large
survey extended from $z=1$ to $z=3$ with an area of $10000\: deg^{2}$
to distinguish with sufficient confidence DGP from $\Lambda$CDM.
In Fig. (\ref{fig-DETF-wp-gamma}) is shown the confidence region
for $w_{0}-\gamma$ for a survey extended from $z=0$ to $z=1.5$
and an area of $20000\: deg^{2}$ (DETF case): the error on $\gamma$
reduces to $\sigma\left(\gamma\right)=0.06$. 

\begin{figure}
\begin{centering}\includegraphics[scale=0.3]{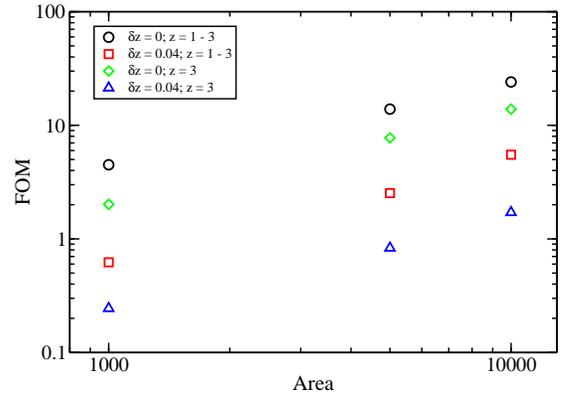}\par\end{centering}

\caption{\label{fig-FOM-w0-w1}Effect of Survey geometry on the dark energy
FOM. We plotter the FOM ($w_{p}-w_{1}$) for spectroscopic surveys
($\delta z=0$) and photometric surveys ($\delta z=0.04$) as a function
of the area.}
\end{figure}

\begin{figure}
\begin{centering}\includegraphics[scale=0.3]{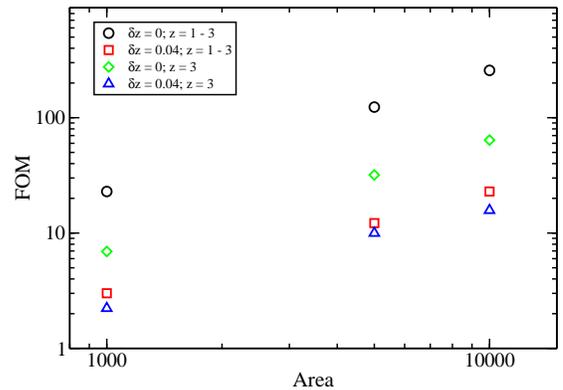}\par\end{centering}

\caption{\label{fig-FOM-w0-gamma}Effect of Survey geometry on the dark energy
FOM. We plotter the FOM ($w_{0}-\gamma$) for spectroscopic surveys
($\delta z=0$) and photometric surveys ($\delta z=0.04$) as a function
of the area.}
\end{figure}

In Fig. (\ref{fig-FOM-w0-w1}) we also show the figure-of-merit (FOM)
suggested by the Dark Energy Task Force report ( Albrecht et al. 2006)
as a simple measure of the constraining power of an experiment. The
FOM is defined as the inverse of the area that encloses the 95\% confidence
region and can be found simply as $(6.17\pi\sqrt{\det F})^{-1}$.
In Fig. (\ref{fig-FOM-w0-gamma}) we plot the FOM for $w_{0}$ and
$\gamma$. The general trend is that the FOM for spectroscopic surveys
are roughly 4-6 times higher than for similar 4\% error photo-$z$
surveys. It will be interesting to compare our FOM on the plane $w_{0},\gamma$
with those obtained from other experiments. This task will be performed
in future work.


\begin{acknowledgments}
D.S. is supported by the Swiss NSF. It is a pleasure to thank Martin
Kunz for interesting discussions.
\end{acknowledgments}

\end{document}